\documentclass[11pt,a4paper]{article}
\usepackage{amsmath}
\usepackage{amsfonts}
\usepackage{diagbox}
\usepackage{array}
\usepackage{tikz}
\usetikzlibrary{fit}					
\usetikzlibrary{backgrounds}	%

\usepackage{mathrsfs}
\usepackage{cite}

\begin{document}

\title{Coefficients at powers of logarithms in the HD+MSL renormalization scheme}

\author{
N.P.Meshcheriakov\,${}^{ab}$, V.V.Shatalova\,${}^{c}$, and K.V.Stepanyantz${}^{abd}$ $\vphantom{\Big(}$
\medskip\\
{\small{\em Moscow State University, Faculty of Physics,}}\\
${}^a${\small{\em Department of Theoretical Physics,}}\\
${}^b${\small{\em Department of Quantum Theory and High Energy Physics,}}\\
{\small{\em 119991, Moscow, Russia}}\\
\vphantom{1}\vspace*{-2mm}\\
${}^c${\small{\em Moscow State University, AESC MSU – Kolmogorov boarding school}},\\
{\small{\em Department of Physics,}}\\
{\small{\em 119192, Moscow, Russia}}\\
\vphantom{1}\vspace*{-2mm}\\
$^d$ {\small{\em Bogoliubov Laboratory of Theoretical Physics, JINR,}}\\
{\small{\em 141980 Dubna, Moscow region, Russia.}}\\
\vphantom{1}\vspace*{-2mm}}

\maketitle

\begin{abstract}
For renormalizable theories with a single coupling constant regularized by higher derivatives we investigate the coefficients at powers of logarithms present in the renormalization constants assuming that divergences are removed by minimal subtractions of logarithms. According to this (HD+MSL) renormalization prescription the renormalization constants include only powers of $\ln\Lambda/\mu$, where $\Lambda$ and $\mu$ are the dimensionful regularization parameter and the renormalization point, respectively. We construct general explicit expressions for arbitrary coefficients at powers of this logarithm present in the coupling constant renormalization and in the field renormalization constant which relate them to the $\beta$-function and (in the latter case) to the corresponding anomalous dimension. To check the correctness, we compare the results with the explicit four-loop calculation made earlier for ${\cal N}=1$ SQED and (for the supersymmetric case) rederive a relation between the renormalization constants following from the NSVZ equation.
\end{abstract}

\unitlength=1cm

\section{Introduction}
\hspace*{\parindent}

Quantum corrections in most field theory models are divergent in the ultraviolet region. In renormalizable theories these divergences can be eliminated by redefining coupling constants, fields, masses, and gauge parameters \cite{Bogolyubov:1980nc}. However, the renormalization procedure admits a certain degree of arbitrariness. The renormalization constants are defined ambiguously because they depend on a regularization and a subtraction scheme. For instance, in the case of using dimensional regularization \cite{'tHooft:1972fi, Bollini:1972ui, Ashmore:1972uj, Cicuta:1972jf} ultraviolet divergences have the form of $\varepsilon$-poles, and one of the simplest renormalization prescriptions (the MS scheme) is to include only pole terms into the renormalization constants \cite{tHooft:1973mfk}. It is important that various divergent contributions to the renormalization constants are not independent \cite{tHooft:1973mfk}. Namely, the coefficients at higher $\varepsilon$-poles in a certain order of the perturbation theory are determined by the coefficients at the lower poles in the previous orders. The corresponding recurrence relations often referred to as the pole equations reveal that all coefficients at the higher order poles can eventually be expressed in terms of the residua of the simple poles. The latter, in turn, are totally determined by the renormalization group functions (RGFs), i.e. by the $\beta$-functions and the anomalous dimensions.

The pole equations have also been generalized to different cases including nonrenormalizable theories \cite{Kazakov:1987jp,Kazakov:2016wrp,Borlakov:2016mwp,Kazakov:2019wce,Solodukhin:2020vuw}. Similar relations can be written for logarithms in the renormalized Green functions \cite{Collins:1984xc}. In the case of using regularizations formulated in $D=4$ \cite{Gnendiger:2017pys} relations between various divergences analogous to those following from the pole equations should also appear, but, to the best of our knowledge, they have been little studied in literature. However, in certain situations the use of the dimensional technique can be inappropriate. For example, dimensional regularization explicitly breaks supersymmetry \cite{Delbourgo:1974az}, while dimensional reduction \cite{Siegel:1979wq} in its mathematically consistent version causes the loss of supersymmetry in higher orders of perturbation theory \cite{Avdeev:1981vf,Avdeev:1982xy,Avdeev:1982np,Velizhanin:2008rw}. On the other hand, an invariant regularization of the supersymmetric theories can be constructed with the help of the higher (covariant) derivative method  \cite{Slavnov:1971aw,Slavnov:1972sq}.  It can be formulated explicitly in terms of the $ \mathcal{N} = 1 $ superfields \cite{Krivoshchekov:1978xg,West:1985jx} and in $\mathcal{N} = 2$ harmonic superspace \cite{Buchbinder:2015eva}. For a long time it was believed that the higher derivative regularization is ill-suited for multiloop calculations, mainly due to significant complication of Feynman rules for vertices.  Despite this, a lot of such calculations (see, e.g., \cite{Shirokov:2022jyd,Kuzmichev:2021lqa,Kuzmichev:2021yjo,Aleshin:2020gec,Kuzmichev:2019ywn,Stepanyantz:2019lyo,Kazantsev:2018kjx,Kazantsev:2018nbl,Shakhmanov:2017soc}) for ${\cal N} = 1$ supersymmetric gauge theories have been made to date. Moreover, the higher covariant derivative regularization appeared to be a very important ingredient of the all-loop perturbative derivation of the NSVZ $\beta$-function \cite{Novikov:1983uc,Jones:1983ip,Novikov:1985rd,Shifman:1986zi} for ${\cal N} = 1$ supersymmetric gauge theories made in \cite{Stepanyantz:2016gtk,Stepanyantz:2019ihw,Stepanyantz:2019lfm,Stepanyantz:2020uke}.
It was shown \cite{Stepanyantz:2020uke} that some all-loop NSVZ schemes (i.e., the ones in which the NSVZ $\beta$-function is valid in all orders) in non-Abelian case are given by the HD+MSL prescription \cite{Kataev:2013eta}. This means that a theory is regularized by higher covariant derivatives and a renormalization procedure is performed with the help of minimal subtractions of logarithms \cite{Shakhmanov:2017wji,Stepanyantz:2017sqg}. Thus, the HD+MSL scheme is an obvious analogue of the MS scheme in the case of using the dimensional technique, but instead of $\varepsilon$-poles the renormalization constants include only powers of $\ln \Lambda/\mu $, where $\Lambda$ is a dimensionful parameter of the higher derivative regularization and $\mu$ is a normalization point. Therefore, it is reasonable to assume that in the HD+MSL scheme it is possible to write the relations between the coefficients at powers of $\ln\Lambda/\mu$ similar to those that follow from the pole equations. Obtaining these relations in the HD+MSL scheme from the renormalization group equations (see, e.g., \cite{Bogolyubov:1980nc,Collins:1984xc}) is the main purpose of this paper. We are going to investigate them in the context of both coupling constant and field renormalizations.

The paper is organized as follows. In Section~\ref{Section_RGE_HD_MSL} we describe the renormalization of the coupling constant and introduce the HD+MSL boundary conditions. The explicit expressions for the coefficients at powers of $\ln\Lambda/\mu$ in the coupling constant renormalization (depending on the coefficients in the perturbative expansion of the $\beta$-function) are derived in Section~\ref{Section_B_nk}. Similar expressions for the coefficients in the field renormalization are obtained in Section~\ref{Section_About_C_nk}. Some examples are considered in Section~\ref{Section_Examples}. In particular, in Subsection~\ref{Subsection:SQED} we compare the general expressions for the coefficients with the known results for ${\cal N}=1$ supersymmetric quantum electrodynamics (SQED) in the four-loop approximation for the coupling constant renormalization and in the three-loop approximation for the matter superfield renormalization. Next, in Subsection~\ref{Subsection:NSVZ_SQED} we derive the equation relating the renormalizations of the coupling constant and of the matter superfields following from the NSVZ equation for ${\cal N}=1$ SQED. The similar investigation for non-Abelian ${\cal N}=1$ supersymmetric gauge theories with a single coupling is made in Subsection~\ref{Subsection:NSVZ_SYM}. In Appendix \ref{Appendix_Powers} we present expressions for the coefficients at powers of $\ln\Lambda/\mu$ which appear in the functions $\ln\alpha/\alpha_0$ and $(\alpha/\alpha_0)^S$ (for an arbitrary real number $S$), where $\alpha$ and $\alpha_0$ are the renormalized and bare coupling constants, respectively.

\section{The coupling constant renormalization and the HD+MSL prescription}
\label{Section_RGE_HD_MSL}
\hspace*{\parindent}

We consider a $4D$ renormalizable gauge theory with a single dimensionless coupling constant $\alpha \equiv e^2/4\pi$, assuming that all divergences are logarithmical and the higher derivative method \cite{Slavnov:1971aw,Slavnov:1972sq} is used for the regularization. This regularization is introduced by adding a term with higher powers of (covariant) derivatives to the Lagrangian. As a result, higher powers of momenta appear in the denominators of the propagators ensuring the (superficial) ultraviolet convergence of the integrals coming from all diagrams beyond the one-loop approximation. According to \cite{Slavnov:1977zf}, the remaining one-loop divergences and subdivergences can be regularized by inserting certain Pauli-Villars determinants into the generating functional. (The form of the Pauli--Villars determinants in the supersymmetric case can be found in \cite{Aleshin:2016yvj,Kazantsev:2017fdc}.) We always assume that the masses of the Pauli-Villars fields $M_I$ are proportional to the parameter $\Lambda$ present in the higher derivative term, and the ratios $M_I/\Lambda$ do not depend on the coupling constant. This implies that the only independent dimensionful parameter $\Lambda$ is present in the regularized theory.

Divergences appearing in quantum corrections are absorbed into the renormalization of coupling constant, fields, and other parameters. In particular, the renormalization of the coupling constant is written as

\begin{equation}
 \frac{1}{\alpha_0} = \frac{Z_{\alpha}}{\alpha},
\end{equation}

\noindent
where the bare coupling constant is marked by the subscript $0$ and $Z_{\alpha} = Z_{\alpha}\left(\alpha, \ln \Lambda/\mu\right)$ is the charge renormalization constant. The relation between the renormalized and bare coupling constants can be presented in the form

\begin{equation}\label{1/a=1/a_0}
\frac{1}{\alpha} = \frac{1}{\alpha_0}+B_{1,0}+B_{1,1} \ln\frac{\Lambda}{\mu}+\sum\limits_{n=2}^\infty \alpha_0^{n-1} \sum\limits_{m=0}^{n-1} B_{n, m} \,  \ln^m \frac{\Lambda}{\mu},
\end{equation}

\noindent
where $B_{n,m}$ are numerical coefficients. Note that these coefficients are not uniquely defined due to their dependence on a subtraction scheme. The HD+MSL renormalization prescription \cite{Kataev:2013eta} can be defined by the boundary conditions

\begin{equation}\label{Boundary_Z_HD+MSL}
Z_i(\alpha_0, \ln \Lambda / \mu = 0) = 1,
\end{equation}

\noindent
where $Z_i$ denotes all renormalization constants of the theory under consideration. In particular, from the boundary condition for $Z_\alpha$ we obtain

\begin{equation}\label{BoundaryAlphaHD+MSL}
\alpha\left(\alpha_0, \ln \Lambda / \mu = 0 \right) = \alpha_0.
\end{equation}

\noindent
This implies that in the HD+MSL scheme

\begin{equation}\label{B_k0=0}
B_{n, 0} = 0, \qquad  n \geq 1.
\end{equation}

Similar to the pole equations valid in the MS scheme in the case of using the dimensional technique, we expect that the coefficients at higher powers of $\ln \Lambda/\mu$ are determined by the ones at the lower powers in the previous orders of perturbation theory and, eventually, by the coefficients at $\ln^{1} \Lambda/\mu$ \cite{Collins:1984xc}. For the considered HD+MSL scheme this is schematically depicted in Fig.~\ref{Fig1}. The coefficients $B_{n+1,n}$ with $n \geq 1$  (of the terms containing $\alpha_0^n \ln^n \Lambda/\mu$) correspond to the leading logarithm approximation and form the main diagonal. Basing on the analogy with the pole equations we expect that for $n \geq 2$ these coefficients can be expressed in terms of $B_{2,1}$ and  $B_{1,1}$. The second diagonal is formed by the coefficients $B_{n+2,n}$ with $n \geq 1$ of the terms containing $\alpha_0^{n+1} \ln^{n} \Lambda/\mu$. For this diagonal the coefficients at higher powers of $\ln\Lambda/\mu$ appear to be determined by $B_{3,1}$,  $B_{2,1}$, and  $B_{1,1}$. In the next section we will show how this can be derived for an arbitrary coefficient of any diagonal and present the corresponding explicit expressions.

\hfill
\begin{figure}[htbp]
	\centering
	
	\tikzstyle{background}=[rectangle,
	fill=gray!20,
	inner sep=0.2cm,
	rounded corners=5mm]

	
	\begin{tikzpicture}[>=latex,text height=1.6ex,text depth=0.3ex]
		
		\matrix[row sep=0.5cm,column sep=0.4cm] {
			&
			\node (ln0)   {$1$};                   &
			\node (ln1)   {$\ln \Lambda/\mu$};     &
			\node (ln2)   {$\ln^2 \Lambda/\mu$};   &
			\node (ln3)   {$\!\ln^3 \Lambda/\mu$}; &
			\node (ln4)   {$\!\ln^4 \Lambda/\mu$}; &
			\node (ln5)   {$\quad\:\dots$};        &
			\\
			\node (a0)    {$1$};              &
			\node (c10)   {$0$};        &
			\node (c11)   {$B_{1,1}$};        &
			\node (c12)   {$ 0 $};            &
			\node (c13)   {$ 0 $};            &
			\node (c14)   {$ 0 $};            &
			\node (c15)   {$\quad\: \dots $}; &
			\\
			\node (a1)    {$\alpha_0$};       &
			\node (c20)   {$0$};        &
			\node (c21)   {$B_{2,1}$};        &
			\node (c22)   {$ 0 $};            &
			\node (c23)   {$ 0 $};            &
			\node (c24)   {$ 0 $};            &
			\node (c25)   {$\quad\: \dots $}; &
			\\
			\node (a2)    {$\alpha_0^2$};     &
			\node (c30)   {$0$};        &
			\node (c31)   {$B_{3,1}$};        &
			\node (c32)   {$B_{3,2} $};      &
			\node (c33)   {$ 0 $};            &
			\node (c34)   {$ 0 $};            &
			\node (c35)   {$\quad\: \dots $}; &
			\\
			\node (a3)    {$\alpha_0^3$};     &
			\node (c40)   {$0$};        &
			\node (c41)   {$B_{4,1}$};        &
			\node (c42)   {$B_{4,2} $};      &
			\node (c43)   {$B_{4,3} $};      &
			\node (c44)   {$ 0 $};            &
			\node (c45)   {$\quad\:\dots $};  &
			\\
			\node (a4)    {$\alpha_0^4$};     &
			\node (c50)   {$0$};        &
			\node (c51)   {$B_{5,1}$};        &
			\node (c52)   {$B_{5,2} $};      &
			\node (c53)   {$B_{5,3} $};      &
			\node (c54)   {$B_{5,4} $};      &
			\node (c55)   {$\quad\:\dots $};  &
			\\
			\node (a5)    {$\hphantom{I}\vdots\hphantom{I}$};      &
			\node (c60)   {$\hphantom{II}\vdots\hphantom{II}$};    &
			\node (c61)   {$\hphantom{II}\vdots\hphantom{II}$};    &
			\node (c62)   {$ \hphantom{II} \vdots\hphantom{II} $}; &
			\node (c63)   {$ \hphantom{II}\vdots \hphantom{II}$};  &
			\node (c64)   {$\hphantom{II}\vdots\hphantom{II} $};   &
			\node (c65)   {$\quad\ddots\! $};                      &
			\\
		};
		
		\path[->]
		(c21) edge (c32)
		(c32) edge (c43)
		(c43) edge (c54)
		(c54) edge (c65)
		
		(c31) edge (c42)
		(c42) edge (c53)
		(c53) edge (c64)
		
		(c41) edge (c52)
		(c52) edge (c63)
		
		(c51) edge (c62)
		;
		
		\begin{pgfonlayer}{background}
			\node [background,
			fit=(ln0) (ln5)] {};
			\node [background,
			fit=(a0) (a5)] {};
		\end{pgfonlayer}
		
	\end{tikzpicture}
	
	\caption{The coefficients $B_{n,m} $ in the coupling constant renormalization (see Eq.~\ref{1/a=1/a_0}) in the HD+MSL scheme. For $m \geq 2$ they are not independent and can be expressed in terms of the coefficients $B_{n,1}$.}
	\label{Fig1}
\end{figure}

\section{Coefficients in the coupling constant renormalization}
\hspace*{\parindent}\label{Section_B_nk}

In this paper we will (standardly) define RGFs in terms of the renormalized coupling constant. In particular, the $\beta$-function is given by the expression

\begin{equation}\label{betaDEFINITION}
\beta(\alpha) \equiv \left.\frac{d\alpha\left(\alpha_0,\ln\Lambda/\mu\right)}{d\ln\mu} \right|_{\alpha_0=\text{const}},
\end{equation}

\noindent
where the derivative is taken at a fixed value of the bare coupling constant. Note that in the HD+MSL scheme RGFs defined in terms of the renormalized couplings coincide with the ones defined in terms of the bare couplings after a formal replacement $\alpha\to\alpha_0$, see \cite{Kataev:2013eta} for detail.

Differentiating Eq.~(\ref{1/a=1/a_0}) with respect to $\ln\mu$ and using the definition (\ref{betaDEFINITION}) we obtain the equation

\begin{eqnarray}\label{BetaThroughB}
\frac{\beta(\alpha)}{\alpha^2} = -\frac{d}{d \ln\mu}\Big(\frac{1}{\alpha}\Big)\bigg|_{\alpha_0=\text{const}}=B_{1,1}+\sum\limits_{n=2}^\infty \alpha_0^{n-1} \sum\limits_{m=1}^{n-1} m \, B_{n, m}  \ln^{m-1}\frac{\Lambda}{\mu}.
\end{eqnarray}

\noindent
Next, we substitute into this equation the standard perturbative expansion for the $\beta$-function

\begin{equation}\label{Beta_PT_Expansion}
\beta(\alpha)=\sum\limits_{n=1}^\infty \beta_n \alpha^{n+1}
\end{equation}

\noindent
and formally set $\Lambda=\mu$. Then, taking into account the boundary condition (\ref{BoundaryAlphaHD+MSL}) it is possible to relate the coefficients of the $\beta$-function to the coefficients of the series (\ref{1/a=1/a_0}) in the HD+MSL scheme,

\begin{equation}\label{Bk1=bk}
\beta_n = B_{n,1}, \qquad n \geq 1.
\end{equation}

To express the remaining coefficients $B_{n,m}$ with $n > m \geq 2$ in terms of $\beta_n$, we substitute the expansion (\ref{Beta_PT_Expansion}) into the left hand side of Eq.~(\ref{BetaThroughB}) and differentiate the resulting equation $(p-1)$ times ($p\ge 2$) with respect to $\ln\mu$ at a fixed value of the bare coupling constant,

\begin{equation}\label{p-1_Differentiations}
\quad \sum\limits_{k=2}^\infty \beta_k\, \frac{d^{p-1} \alpha^{k-1}}{d \ln\mu^{p-1}}= (-1)^{p-1} \sum\limits_{n=2}^\infty  \alpha_0^{n-1}  \sum\limits_{m=1}^{n-1} m(m-1)\dots(m-p+1) \, \theta [m-p] \, B_{n,m}  \ln^{m-p}\frac{\Lambda}{\mu}.\quad
\end{equation}

\noindent
Here

\begin{equation}\label{Heaviside}
	\theta[m-p] \equiv
	\begin{cases}
		0, \quad m<p,\\	
		1, \quad p\geq m,
	\end{cases}
\end{equation}

\noindent
is the Heaviside step function, and we took into account that the coefficients $\beta_k$ do not depend on  $\ln\Lambda/\mu$. It is convenient to represent the derivatives in the left hand side of Eq.~(\ref{p-1_Differentiations}) using the chain rule,

\begin{equation}\label{Chain_rule}
\frac{d}{d \ln\mu}\bigg|_{\alpha_0=\text{const}} = \frac{d\alpha}{d\ln\mu}\cdot \frac{d}{d \alpha} = \beta(\alpha)\frac{d}{d \alpha} = \sum\limits_{n=1}^\infty \beta_n \alpha^{n+1} \frac{d}{d \alpha}.
\end{equation}

\noindent
With the help of this equation we obtain

\begin{equation}
\frac{d^{p-1} \alpha^{k-1}}{d \ln\mu^{p-1}} = \bigg(\sum\limits_{k_{p-1}=1}^\infty \beta_{k_{p-1}} \alpha^{k_{p-1}+1} \frac{ d}{d \alpha}\bigg) \bigg(\sum\limits_{k_{p-2}=1}^\infty \beta_{k_{p-2}} \alpha^{k_{p-2}+1}
\frac{ d}{d \alpha}\bigg) \dots \bigg(\sum\limits_{k_1=1}^\infty \beta_{k_1} \alpha^{k_1+1} \frac{d}{d \alpha} \alpha^{k-1}\bigg),
\end{equation}

\noindent
where the derivatives with respect to $ \alpha $ act on everything that stands to the right of them. After calculating these derivatives the considered expression can be written as

\begin{eqnarray}\label{MainDifferentiations}	
&& \frac{d^{p-1} \alpha^{k-1}}{d \ln\mu^{p-1}} = \sum\limits_{k_1=1}^\infty (k-1)\beta_{k_1} \sum\limits_{k_2=1}^\infty (k-1+k_1) \beta_{k_2} \sum\limits_{k_3=1}^\infty (k-1+k_1+k_2)\beta_{k_3} \dots \nonumber\\
&&\qquad\qquad\qquad\qquad \times \sum\limits_{k_{p-1}=1}^\infty (k-1+k_1+k_2+\dots+k_{p-2}) \beta_{k_{p-1}} \alpha^{k-1+k_1+k_2+\dots+k_{p-1}}.\qquad
\end{eqnarray}

Next, we set $ \Lambda = \mu $ in Eq.~(\ref{p-1_Differentiations}) and use the HD+MSL boundary condition (\ref{BoundaryAlphaHD+MSL}). Then only terms with $m=p$ in the right hand side of Eq.~(\ref{p-1_Differentiations}) survive, so that

\begin{eqnarray}\label{FormalRelations}
&& \sum\limits_{k=2}^\infty \beta_k  \sum\limits_{k_1=1}^\infty (k-1)\beta_{k_1} \sum\limits_{k_2=1}^\infty (k-1+k_1) \beta_{k_2} \sum\limits_{k_3=1}^\infty (k-1+k_1+k_2)\beta_{k_3} \dots
\sum\limits_{k_{p-1}=1}^\infty (k-1 \qquad\nonumber\\
&& +k_1+k_2+\dots+k_{p-2}) \beta_{k_{p-1}} \alpha_0^{k-1+k_1+k_2+\dots+k_{p-1}}
= (-1)^{p-1}p!\sum\limits_{n=1}^\infty \alpha_0^{n+p-1} B_{n+p,\,p}.\qquad
\end{eqnarray}

\noindent
Equating the coefficients at the same powers of $\alpha_0$ in Eq.~(\ref{FormalRelations}) we obtain the explicit expression for $B_{n+p,\,p}$ with $n\ge 1$ and $p\ge 2$,

\begin{eqnarray}\label{GeneralAnaliticalResult}
&& B_{n+p,\,p} = (-1)^{p-1}\frac{1}{p!}\,\overset{n+1}{\underset{k=2}{\sum}}\, \beta_k \sum_{k_1} (k-1) \beta_{k_1} \sum_{k_2}(k-1+k_1) \beta_{k_2}\sum_{k_3} (k-1+k_1+k_2)\beta_{k_3}\dots
\qquad\nonumber\\
&& \times \sum_{k_{p-1}}(k-1+k_1+k_2+\dots+k_{p-2})\beta_{k_{p-1}}\bigg|_{\substack{k_1+k_2+\dots +k_{p-1}=n+p-k}},
\end{eqnarray}

\noindent
where the sums over the indices $k_i$ are taken only over the  positive integers satisfying the condition $ k_1+k_2+\dots +k_{p-1}=n+p-k $. Besides, we set the upper limit of summation over $ k $ equal to $ n+1 $ since the higher values obviously do not meet this condition.

For the lowest values of $n$ (corresponding to the main and two subleading diagonals) the coefficients (\ref{GeneralAnaliticalResult}) can be presented in a different form,

\begin{eqnarray}\label{B_p+1_p}
&& B_{1+p, \, p} = (-1)^{p-1} \frac{1}{p}\,\beta_1^{p-1}\beta_2; \\
\label{B_p+2_p}	
&& B_{2+p, \, p} = (-1)^{p-1} \left( \beta_1^{p-1}\beta_3 +  (H_p-1)\theta[p-2]\,\beta_1^{p-2} \beta_2^2 \right); \\
\label{B_p+3_p}
&& B_{3+p,\, p} = (-1)^{p-1} \frac{(p+1)}{2}\bigg\{ \beta_1^{p-1}\beta_4 + 2  (H_p -1) \theta[p-2]\,\beta_1^{p-2}\beta_2\beta_3 \nonumber\\
&& + \Big(\, \frac{p+3}{p+1}+H_{p+1}( H_{p+1} - 2)-H_{p+1,\, 2}\Big) \theta[p-3]\,\beta_1^{p-3}\beta_2^3 \bigg\},
\end{eqnarray}

\noindent
where the Heaviside step function is given by Eq.~(\ref{Heaviside}), and the generalized harmonic numbers are defined as

\begin{equation}\label{Harmonic_Numbers}
H_{p, \, s} \equiv \sum_{k=1}^{p} \frac{1}{k^s}.
\end{equation}

\noindent
For $s = 1$ they correspond to the usual harmonic numbers $H_p  \equiv  H_{p,1}$. Some particular values of the generalized harmonic numbers are presented in Table~\ref{Table:1}.

\begin{table}[h]
\begin{center}
\begin{tabular}{|c||c|c|c|c|c|c|c|c|}
	\hline
	\diagbox{$ s $}{$ p $}  &\hspace*{5mm} 1\hspace*{5mm} &\hspace*{5mm} 2\hspace*{5mm} &\hspace*{5mm} 3\hspace*{5mm} &\hspace*{5mm} 4\hspace*{5mm} &\hspace*{5mm} 5\hspace*{5mm} &\hspace*{5mm} 6\hspace*{5mm} &\hspace*{5mm} 7\hspace*{5mm} &\hspace*{3.5mm} $\dots$\hspace*{3.5mm} \\
	\hline \hline
	1$\vphantom{\Bigg(}$  & 1  & $\ \ {\displaystyle \frac{3}{2}}$\ \ & ${\displaystyle \frac{11}{6}}$ & ${\displaystyle \frac{25}{12}}$ & ${\displaystyle \frac{137}{60}}$  & ${\displaystyle \frac{49}{20}}$ & ${\displaystyle \frac{363}{140}}$ & $\dots$\\[2mm]
	\hline
	2$\vphantom{\Bigg(}$ & 1 & ${\displaystyle \frac{5}{4}}$ & ${\displaystyle \frac{49}{36}}$ & ${\displaystyle \frac{205}{144}}$ & ${\displaystyle \frac{5269}{3600}}$ & ${\displaystyle \frac{5369}{3600}}$ &
${\displaystyle \frac{266681}{176400}}$ & $\dots$\\[2mm]
	\hline
	3$\vphantom{\Bigg(}$ & 1 & ${\displaystyle \frac{9}{8}}$ & ${\displaystyle \frac{251}{216}}$ & ${\displaystyle \frac{2035}{1728}}$ & ${\displaystyle \frac{256103}{216000}}$ & ${\displaystyle \frac{28567}{24000}}$ &
${\displaystyle \frac{9822481}{8232000}}$ & $\dots$\\[2mm]
	\hline
    $\vdots$$\vphantom{\Bigg(}$ & $\vdots$ & $\vdots$ & $\vdots$ & $\vdots$ & $\vdots$ & $\vdots$ & $\vdots$ & $\ddots$\\[2mm]
    \hline
\end{tabular}
\end{center}
\caption{Some lower values of the generalized harmonic numbers $ H_{p, \, s} $.}
\label{Table:1}
\end{table}

\noindent
Then, for the lowest coefficients in the 2-nd and 3-rd subdiagonals in the coupling constant renormalization we get

\begin{align}\label{B_Explicit}
&B_{4,2}=-\beta_1\beta_3-\frac{1}{2}\beta_2^2; \quad &B_{5,2}&=-\frac{3}{2}\beta_1\beta_4 -\frac{3}{2}\beta_2\beta_3;\nonumber\\
&B_{5,3}=\,\beta_1^2\beta_3 + \frac{5}{6}\beta_1\beta_2^2; \quad
&B_{6,3}&=2\beta_1^2\beta_4+\frac{10}{3}\beta_1\beta_2\beta_3+ \frac{1}{2}\beta_2^3;\nonumber\\
&B_{6,4}=-\beta_1^3\beta_3 -\frac{13}{12}\beta_1^2\beta_2^2;\quad &B_{7,4}&=-\frac{5}{2}\beta_1^3\beta_4-\frac{65}{12}\beta_1^2\beta_2\beta_3 -\frac{35}{24}\beta_1\beta_2^3.
\end{align}

In Appendix~\ref{Appendix_Powers} using the method described in this section we find the coefficients at powers of $\ln\Lambda/\mu$ in the expressions $\ln\alpha/\alpha_0$ and $(\alpha/\alpha_0)^S$ (for an arbitrary real number $S$). The results are given by Eqs.~(\ref{Ln_Coefficients}) and (\ref{AlphsS_Coefficients}), respectively.

\section{Coefficients in the field renormalization}
\hspace*{\parindent}\label{Section_About_C_nk}

Similarly, it is possible to find explicit expressions for the coefficients in the field renormalization constants. Let us denote such a constant by $Z$. Then we can present $\ln Z$ as the series

\begin{equation}\label{lnZ_via_Cnk}
\ln Z\Big(\alpha\Big(\alpha_0,\ln\frac{\Lambda}{\mu}\Big),\ln\frac{\Lambda}{\mu}\Big) = \sum_{n=1}^{\infty} \alpha_0^n \sum_{m=0}^{n} C_{n,m} \ln^m \frac{\Lambda}{\mu}.
\end{equation}

\noindent
The corresponding anomalous dimension is defined in terms of the renormalized coupling constant by the equation

\begin{equation}\label{Gamma_Definition}
\gamma(\alpha) \equiv \left. \frac{d \ln Z }{d \ln\mu} \right|_{\alpha_0=\text{const}}
\end{equation}

\noindent
and has the perturbative expansion

\begin{equation}\label{gamma_perturbative}
\gamma(\alpha) = \sum_{n=1}^{\infty} \gamma_n \alpha^n,
\end{equation}

\noindent
where the coefficients $\gamma_n$ do not contain $\ln\Lambda/\mu$. From the other side, substituting the expression (\ref{lnZ_via_Cnk}) into Eq.~(\ref{Gamma_Definition}) we obtain the anomalous dimension as a function of the bare coupling constant,

\begin{eqnarray}\label{gamma_via_C}
\gamma\Big(\alpha\Big(\alpha_0,\ln\frac{\Lambda}{\mu}\Big)\Big) = - \sum_{n=1}^{\infty} \alpha_0^n \sum_{m=1}^{n} m\, C_{n,m} \ln^{m-1} \frac{\Lambda}{\mu}.
\end{eqnarray}

\noindent
Next, we equate Eqs.~(\ref{gamma_perturbative}) and (\ref{gamma_via_C}) and repeat the reasoning of Section \ref{Section_B_nk}. Namely, it is necessary to differentiate the resulting equation $(p-1)$ times with respect to $\ln\mu$, set $\mu=\Lambda$, and apply the HD+MSL boundary conditions (\ref{Boundary_Z_HD+MSL}) and (\ref{BoundaryAlphaHD+MSL}). Equating the coefficients of the terms with the same powers of $\alpha_0$ we obtain explicit expressions for the coefficients $C_{n,m}$ in the HD+MSL scheme,

\begin{eqnarray}\label{Bn1=-gamma_n}
&& C_{n,0} = 0;\qquad C_{n,1} = - \gamma_n,\qquad n\ge 1;\vphantom{\frac{1}{2}}\\
\label{GeneralResult_FieldRen}
&& C_{n+p,\, p} = (-1)^{p} \frac{1}{p!} \sum_{k=1}^{n+1} \gamma_k \sum_{k_1}  k \beta_{k_1} \sum_{k_2} (k+k_1) \beta_{k_2} \sum_{k_3}(k+k_1+k_2) \beta_{k_3}\dots \nonumber\\
&& \times \sum_{k_{p-1}} (k+k_1+k_2+\dots+k_{p-2})\beta_{k_{p-1}}\bigg|_{\substack{k_1+k_2+\dots +k_{p-1}=n+p-k}}, \qquad n\ge 0,\quad p\ge 2.\qquad
\end{eqnarray}

\noindent
Their structure is very similar to Eq.~(\ref{GeneralAnaliticalResult}) except for the presence of the coefficients $ \gamma_k $ instead of $\beta_k$ in the first sum and slightly different numerical factors. Actually, making the formal replacement

\begin{equation}\label{Replacement}
\gamma_k \to - \beta_{k+1} 	
\end{equation}

\noindent
in the first sum of Eq.~(\ref{GeneralResult_FieldRen}) one can obtain the coefficient $B_{n+1+p,\,p} $ from $C_{n+p,\,p} $.

Using Eq.~(\ref{GeneralResult_FieldRen}) it is possible to derive explicit expressions for all coefficients in certain diagonals. In particular, the coefficients corresponding to the main and two subleading diagonals can be presented in the form

\begin{eqnarray}\label{C_p_p}
&& C_{p, \, p} = \frac{(-1)^{p}}{p}\,  \beta_1^{p-1} \gamma_1;\\
\label{C_p+1_p}
&& C_{1+p, \, p} = (-1)^{p} \left( \beta_1^{p-1} \gamma_2  +  (H_p-1)\theta[p-2]\, \beta_1^{p-2} \beta_2 \gamma_1 \right); \\
\label{C_p+2_p}
&& C_{2+p,\, p} = (-1)^{p} \frac{(p+1)}{2}\bigg\{ \beta_1^{p-1} \gamma_3  + \theta[p-2]\,\beta_1^{p-2} \Big( (2 H_{p+1} -3) \beta_2 \gamma_2 + \frac{(p-1)}{(p+1)} \beta_3 \gamma_1 \Big) \nonumber\\
&& + \Big(\, \frac{p+3}{p+1}+H_{p+1}( H_{p+1} - 2)-H_{p+1,\, 2}\Big) \theta[p-3]\, \beta_1^{p-3}\beta_2^2 \gamma_1 \bigg\},	
\end{eqnarray}

\noindent
where the generalized harmonic numbers $H_{p, \, s}$ are defined by Eq.~(\ref{Harmonic_Numbers}). In particular, from Eqs.~(\ref{C_p+1_p}) and (\ref{C_p+2_p}) we obtain that the lowest coefficients in the 2-nd and 3-rd subdiagonals are written as

\begin{align}\label{C_Explicit}
&C_{3,2}=\beta_1 \gamma_2 +\frac{1}{2} \beta_2 \gamma_1; \quad
&C_{4,2}&=\frac{3}{2} \beta_1 \gamma_3 + \beta_2 \gamma_2 + \frac{1}{2} \beta_3 \gamma_1; \nonumber\\
&C_{4,3}= - \beta_1^2 \gamma_2 - \frac{5}{6} \beta_1 \beta_2 \gamma_1; \quad
&C_{5,3}&=-2 \beta_1^2 \gamma_3 - \frac{7}{3} \beta_1 \beta_2 \gamma_2 - \beta_1 \beta_3 \gamma_1 - \frac{1}{2} \beta_2^2 \gamma_1;\nonumber\\
&C_{5,4}= \beta_1^3 \gamma_2 + \frac{13}{12} \beta_1^2 \beta_2 \gamma_1; \quad
&C_{6,4}&= \frac{5}{2} \beta_1^3 \gamma_3 + \frac{47}{12} \beta_1^2 \beta_2 \gamma_2 + \frac{3}{2} \beta_1^2 \beta_3 \gamma_1  + \frac{35}{24} \beta_1 \beta_2^2 \gamma_1.
\end{align}

\section{Examples}
\label{Section_Examples}

\subsection{Renormalization of ${\cal N}=1$ SQED with $N_f$ flavors}
\hspace*{\parindent}\label{Subsection:SQED}

Let us compare the general expressions obtained in the previous sections with the results of explicit multiloop calculations made for ${\cal N}=1$ SQED with $N_f$ flavors regularized by higher derivatives. The three-loop anomalous dimension of the matter superfields for this theory was found in \cite{Shirokov:2022jyd}. After that, the four-loop $\beta$-function was constructed with the help of the NSVZ equation. The results in the HD+MSL scheme are given by the equations

\begin{eqnarray}\label{Beta_4loop_SQED}
&& \frac{\beta(\alpha)}{\alpha^2} = \frac{N_f}{\pi}+ \frac{\alpha N_f }{\pi^2} - \frac{\alpha^2}{\pi^3}\Big[\, \frac{N_f}{2} + (N_f)^2 \Big(\ln a +1+\frac{A_1}{2}\Big)\Big] + \frac{\alpha^3}{\pi^4}\Big[\, \frac{N_f}{2}\nonumber\\
&&\qquad\quad + (N_f)^2 \Big(\ln a + \frac{3}{4}+C\Big)
+ (N_f)^3\Big((\ln a +1)^2 -\frac{A_2}{4} + D_1\ln a + D_2 \Big)\Big]  + \mathcal{O}(\alpha^4); \qquad\quad\\
&&\vphantom{1}\nonumber\\
\label{Gamma_3loop_SQED}	
&&\gamma(\alpha) = 	- \frac{\alpha}{\pi}  + \frac{\alpha^2}{\pi^2} \Big[\, \frac{1}{2} + N_f\Big(\ln a +1+\frac{A_1}{2}\Big)\Big]
- \frac{\alpha^3}{\pi^3} \Big[\, \frac{1}{2} + N_f \Big(\ln a + \frac{3}{4}+C\Big)\nonumber\\
&&\qquad\qquad\qquad\qquad\qquad\qquad\quad\ \, + (N_f)^2\Big((\ln a +1)^2 -\frac{A_2}{4} + D_1\ln a + D_2 \Big)\Big] + \mathcal{O}(\alpha^4),\qquad
\end{eqnarray}

\noindent
where $a$, $A_1$, $A_2$, $C$, $D_1$, and $D_2$ are some parameters which depend on a particular version of the higher covariant derivative regularization. Explicit expressions for them can be found in \cite{Shirokov:2022jyd}. For example, $a=M/\Lambda$, where $M$ is the mass of the Pauli--Villars superfield (needed for removing residual one-loop divergences).

The equations describing the coupling constant and matter superfield renormalizations in the HD+MSL scheme in the considered case are written as

\begin{eqnarray}\label{Ren_alpha_4loop}
&&\hspace*{-5mm} \frac{1}{\alpha} = 	\frac{1}{\alpha_0} + \frac{N_f}{\pi} \ln \frac{\Lambda}{\mu} + \frac{\alpha_0 N_f}{\pi^2}  \ln \frac{\Lambda}{\mu} -  \frac{\alpha_0^2 N_f}{\pi^3} \Big[\,\frac{1}{2}\ln\frac{\Lambda}{\mu}
+ N_f\Big(\ln a + 1 + \frac{A_1}{2}\Big) \ln\frac{\Lambda}{\mu} + \frac{N_f}{2}  \ln^2\frac{\Lambda}{\mu} \Big]
\nonumber\\
&&\hspace*{-5mm} + \frac{\alpha_0^3 N_f}{\pi^4} \Big[\, \frac{1}{2} \ln\frac{\Lambda}{\mu} + N_f \Big(\ln a + \frac{3}{4} + C\Big) \ln\frac{\Lambda}{\mu} + (N_f)^2\Big((\ln a +1)^2 -\frac{A_2}{4} + D_1\ln a + D_2 \Big) \ln\frac{\Lambda}{\mu}   \nonumber\\
&&\hspace*{-5mm} + (N_f)^2 \Big(\ln a + 1 +\frac{A_1}{2}\Big) \ln^2\frac{\Lambda}{\mu}
+ \frac{(N_f)^2}{3}  \ln^3 \frac{\Lambda}{\mu}   \Big] + \mathcal{O}(\alpha_0^4);\\
&&\vphantom{1}\nonumber\\
\label{Ren_lnZ_3loop}
&&\hspace*{-5mm} \ln Z = \frac{\alpha_0}{\pi} \ln\frac{\Lambda}{\mu} - \frac{\alpha_0^2}{\pi^2}\, \Big[\, \frac{1}{2} \ln\frac{\Lambda}{\mu} + N_f\Big(\ln a + 1 + \frac{A_1}{2}\Big) \ln\frac{\Lambda}{\mu}  + \frac{N_f}{2} \ln^2 \frac{\Lambda}{\mu} \Big] + \frac{\alpha_0^3}{\pi^3}\, \Big[\, \frac{1}{2} \ln \frac{\Lambda}{\mu} + N_f \quad\nonumber\\
&&\hspace*{-5mm} \times \Big(\ln a + \frac{3}{4} + C\Big) \ln\frac{\Lambda}{\mu}
+ (N_f)^2\Big((\ln a +1)^2 -\frac{A_2}{4} + D_1\ln a + D_2 \Big)  \ln \frac{\Lambda}{\mu} + (N_f)^2 \Big(\ln a +1\nonumber\\
&&\hspace*{-5mm} + \frac{A_1}{2}\Big)  \ln^2 \frac{\Lambda}{\mu} + \frac{(N_f)^2}{3}  \ln^3 \frac{\Lambda}{\mu}\Big]  + \mathcal{O}(\alpha_0^4).
\end{eqnarray}

From Eqs.~(\ref{Beta_4loop_SQED}) --- (\ref{Ren_lnZ_3loop}) we obtain that in the HD+MSL scheme the coefficients of the $\beta$-function and of the anomalous dimension are given by the expressions

\begin{eqnarray}\label{Beta_Lowest}
&& \beta_1 = B_{1,1} = \frac{N_f}{\pi};\qquad\nonumber\\
&& \beta_2 = B_{2,1} = \frac{N_f}{\pi^2};\qquad\nonumber\\
&& \beta_3 = B_{3,1} =  -  \frac{N_f}{\pi^3} \Big[\,\frac{1}{2} + N_f\Big(\ln a + 1 + \frac{A_1}{2}\Big) \Big];\qquad\nonumber\\
&& \beta_4 = B_{4,1} = \frac{N_f}{\pi^4} \Big[\, \frac{1}{2} + N_f \Big(\ln a + \frac{3}{4} + C\Big) + (N_f)^2\Big((\ln a +1)^2 -\frac{A_2}{4} + D_1\ln a + D_2 \Big)\Big];\nonumber\\
&&\vphantom{1}\\
\label{Gamma_Lowest}
&& \gamma_1 = - C_{1,1} = -\frac{1}{\pi};\qquad\nonumber\\
&& \gamma_2 = - C_{2,1} = \frac{1}{\pi^2} \Big[\, \frac{1}{2} + N_f\Big(\ln a +1+\frac{A_1}{2}\Big)\Big];\qquad \nonumber\\
&& \gamma_3 = - C_{3,1} = - \frac{1}{\pi^3} \Big[\, \frac{1}{2} + N_f \Big(\ln a + \frac{3}{4}+C\Big) + (N_f)^2\Big((\ln a +1)^2 -\frac{A_2}{4} + D_1\ln a + D_2 \Big)\Big].\nonumber\\
\end{eqnarray}

\noindent
Substituting them into Eq.~(\ref{GeneralAnaliticalResult}) (or, equivalently, into the relevant equations in (\ref{B_Explicit})) we obtain the expressions for the coefficients at higher powers of $\ln\Lambda/\mu$ in the coupling constant renormalization,

\begin{eqnarray}
&& B_{3,2} = -\frac{1}{2}\beta_1\beta_2= - \frac{(N_f)^2}{2\pi^3}; \\
&& B_{4,2} = -\beta_1\beta_3-\frac{1}{2}\beta_2^2 = \frac{(N_f)^3}{\pi^4}\Big(\ln a + 1 + \frac{A_1}{2}\Big); \\
&& B_{4,3} = \frac{1}{3}\beta_1^2 \beta_2 = \frac{(N_f)^3}{3\pi^4}.
\end{eqnarray}

\noindent
The coincidence of these values with the coefficients in Eq.~(\ref{Ren_alpha_4loop}) confirms the correctness of the general equations presented above.

Similarly, substituting Eqs.~(\ref{Beta_Lowest}) and (\ref{Gamma_Lowest}) into Eq.~(\ref{GeneralResult_FieldRen}) or Eq.~(\ref{C_Explicit}) we find the coefficients at higher powers of $\ln\Lambda/\mu$ in $\ln Z$,

\begin{eqnarray}
&& C_{2,2} = \frac{1}{2}\beta_1\gamma_1= - \frac{N_f}{2\pi^2}; \\
&& C_{3,2} = \beta_1 \gamma_2 +\frac{1}{2} \beta_2 \gamma_1 = \frac{(N_f)^2}{\pi^3}\Big(\ln a + 1 + \frac{A_1}{2}\Big);	\\
&& C_{3,3} = - \frac{1}{3}\beta_1^2 \gamma_1 = \frac{(N_f)^2}{3\pi^3}.
\end{eqnarray}

\noindent
Again, these values are exactly the same as the coefficients in Eq.~(\ref{Ren_lnZ_3loop}), so that the general equations derived earlier really produce the correct result for the theory under consideration.

\subsection{Constraints on the renormalization constants following from the NSVZ $\beta$-function for ${\cal N}=1$ SQED with $N_f$ flavors}
\hspace*{\parindent}\label{Subsection:NSVZ_SQED}

As another example, we consider ${\cal N}=1$ SQED with $N_f$ flavors and check a relation between the renormalizations of the coupling constant and of the matter superfields following from the exact NSVZ $\beta$-function \cite{Vainshtein:1986ja,Shifman:1985fi}

\begin{eqnarray}\label{NSVZ:SQED_renormalized}
\frac{\beta(\alpha)}{\alpha^2}	 = \frac{ N_f}{\pi}\Big(1-\gamma(\alpha)\Big).
\end{eqnarray}

\noindent
According to \cite{Kataev:2013eta}, for the considered theory in the HD+MSL scheme the NSVZ equation is valid in all orders, although for an arbitrary renormalization prescription this is not in general satisfied. Using Eqs.~(\ref{Beta_PT_Expansion}) and (\ref{gamma_perturbative}) we see that in the Abelian case due to the NSVZ equation the $\beta$-function in a certain loop is related to the anomalous dimension in the previous loop,

\begin{eqnarray}\label{SQED_NSVZ_bn+1_gamma_n}
\beta_{n+1} = - \frac{N_f}{\pi} \gamma_n.
\end{eqnarray}

\noindent
Substituting the expression for $\gamma_n$ following from this equation into Eq.~(\ref{GeneralResult_FieldRen}) and comparing the result with Eq.~(\ref{GeneralAnaliticalResult}) we see that the coefficients at higher powers of $\ln\Lambda/\mu$ satisfy the constraints

\begin{equation}
B_{n+p+1, \, p} = \frac{N_f}{\pi} C_{n+p, \, p}, \qquad n\geq 0, \quad p\geq 2.
\end{equation}

\noindent
Taking into account that $B_{1,1} = N_f/\pi$ and using the HD+MSL boundary conditions (\ref{Boundary_Z_HD+MSL}) and (\ref{BoundaryAlphaHD+MSL}) we obtain the relation

\begin{eqnarray}
&& 0 = \Big(B_{1,1} - \frac{N_f}{\pi}\Big)\ln\frac{\Lambda}{\mu} + \sum_{n=1}^{\infty} \alpha_0^n \sum_{m=1}^{n} \Big(B_{n+1, m} -\frac{N_f}{\pi}C_{n,m}\Big) \ln^{m}\Lambda/\mu\nonumber\\
&&\qquad\qquad\qquad\qquad\qquad\qquad\qquad\qquad\qquad\qquad = \frac{1}{\alpha} - \frac{1}{\alpha_0} - \frac{N_f}{\pi} \ln\frac{\Lambda}{\mu} - \frac{N_f}{\pi}\ln Z.\qquad
\end{eqnarray}

\noindent
Exactly this equation is obtained after integrating the NSVZ $\beta$-function (\ref{NSVZ:SQED_renormalized}) with respect to $\ln\mu$, again, taking into account Eqs.~(\ref{Boundary_Z_HD+MSL}) and (\ref{BoundaryAlphaHD+MSL}).
Therefore, Eqs.~(\ref{GeneralAnaliticalResult}) and (\ref{GeneralResult_FieldRen}) have been verified by a nontrivial all-order calculation.

\subsection{Constraints on the renormalization constants following from the NSVZ $\beta$-function for non-Abelian ${\cal N}=1$ supersymmetric theories}
\hspace*{\parindent}\label{Subsection:NSVZ_SYM}

In the non-Abelian case the NSVZ $\beta$-function \cite{Novikov:1983uc,Jones:1983ip,Novikov:1985rd,Shifman:1986zi} for ${\cal N}=1$ supersymmetric theories with a simple gauge group $G$ and the chiral matter superfields in the representation $R$ is given by the expression\footnote{The NSVZ equations for theories with multiple gauge couplings can be found in \cite{Korneev:2021zdz}.}

\begin{eqnarray}\label{NSVZ:Def_renormalized_RGF}
\frac{\beta(\alpha,\lambda)}{\alpha^2} = - \frac{3 C_2 - T(R) + C(R)_i{}^j \gamma_j{}^i (\alpha,\lambda)  /r }{2\pi(1- C_2\alpha/2\pi)},
\end{eqnarray}

\noindent
where

\begin{eqnarray}
&& \mbox{tr}(T^A T^B) \equiv T(R)\,\delta^{AB};\qquad  (T^A T^A)_i{}^j \equiv C(R)_i{}^j; \qquad\vphantom{\Big(}\nonumber\\
&& f^{ACD} f^{BCD} \equiv C_2\, \delta^{AB}; \qquad\quad r \equiv \mbox{dim} \, G = \delta^{AA}.\vphantom{\Big(}
\end{eqnarray}

\noindent
According to \cite{Stepanyantz:2020uke}, it is also valid in all-loops in the HD+MSL scheme. Note that, in general, for non-Abelian theories RGFs can also depend on Yukawa couplings $\lambda^{ijk}$ abbreviated with $\lambda$. However, in this paper we consider theories with a single coupling constant, so that below we will omit the dependence on $\lambda$. Then we rewrite the NSVZ equation in the form

\begin{eqnarray}\label{SYM_NSVZ_lambda=0}
\frac{\beta(\alpha)}{\alpha^2} + \frac{1}{2 \pi } \Big(3 C_2 - T(R) \Big) + \frac{1}{2 \pi r} C(R)_i{}^j \gamma_j{}^i(\alpha) = \frac{C_2}{2 \pi} \frac{\beta(\alpha)}{\alpha}
\end{eqnarray}

\noindent
and substitute the expansions (\ref{Beta_PT_Expansion}) and (\ref{gamma_perturbative}) into it. Equating the coefficients at the same powers of the coupling constant $\alpha$ gives the equations

\begin{eqnarray}\label{Recurrence_relation_NSVZ}
&& \beta_1 = - \frac{1}{2 \pi } \Big(3 C_2 - T(R) \Big); \nonumber\\
&& \beta_{n+1} = \frac{C_2}{2 \pi} \beta_{n} - \frac{1}{2 \pi r} C(R)_i{}^j \left(\gamma_{n} \right)_j{}^i, \qquad n \geq 1.
\end{eqnarray}

\noindent
The solution of the recurrence relations (\ref{Recurrence_relation_NSVZ}) is written as

\begin{equation}
\beta_{n+1} = \Big(\frac{C_2}{2 \pi}\Big)^{n} \beta_{1} - \frac{1}{2 \pi r} C(R)_i{}^j \sum_{k=0}^{n-1} \Big(\frac{C_2}{2 \pi}\Big)^{k}\big(\gamma_{n-k} \big)_j{}^i.
\end{equation}

\noindent
With the help of Eq.~(\ref{Recurrence_relation_NSVZ}) and the expressions (\ref{GeneralAnaliticalResult}), (\ref{GeneralResult_FieldRen}), and (\ref{Ln_Coefficients}) we obtain the equation relating coefficients in the expansions of $1/\alpha$, $\ln\alpha/\alpha_0$, and $(\ln Z)_j{}^i$,

\begin{equation}
B_{n+1+p,\,p} = - \frac{C_2}{2\pi} \widetilde B_{n+p,\,p} + \frac{1}{2\pi r} C(R)_i{}^j (C_{n+p,\,p})_j{}^i,\qquad n\ge 0,\quad p\ge 1.
\end{equation}

\noindent
Taking into account that $B_{1,1} = -(3C_2-T(R))/2\pi$ and using Eqs.~(\ref{1/a=1/a_0}), (\ref{lnZ_via_Cnk}), and (\ref{Ln_Expansion}) this relation can be presented as

\begin{eqnarray}
\frac{1}{\alpha} - \frac{1}{\alpha_0} + \frac{1}{2\pi} \left(3 C_2 - T(R)\right)\ln\frac{\Lambda}{\mu} - \frac{1}{2\pi r} C(R)_i{}^j (\ln Z)_j{}^i + \frac{C_2}{2\pi} \ln\frac{\alpha}{\alpha_0} = 0.
\end{eqnarray}

\noindent
It exactly coincides with the equation obtained after integrating Eq.~(\ref{SYM_NSVZ_lambda=0}) with respect to $\ln\mu$ and taking into account the HD+MSL boundary conditions (\ref{Boundary_Z_HD+MSL}) and (\ref{BoundaryAlphaHD+MSL}). This fact can be considered as a very nontrivial correctness test of the calculations made in this paper.

\section{Conclusion}
\hspace*{\parindent}

In this paper we present explicit expressions for the coefficients at powers of $\ln\Lambda/\mu$ in the renormalization of the coupling constant and fields for renormalizable theories with a single coupling in the HD+MSL scheme
(assuming that the corresponding divergences are logarithmical). In the HD+MSL scheme a theory is regularized by higher covariant derivatives, and divergences are removed by minimal subtractions of logarithms, when only $(\ln\Lambda/\mu)^k$ with $k\ge 1$ are included into the renormalization constants. This subtraction scheme is similar to minimal subtraction in the case of using the dimensional technique, and the expressions obtained in this paper (Eqs.~(\ref{GeneralAnaliticalResult}) and (\ref{GeneralResult_FieldRen})) are analogous to the solutions of the pole equations. They express coefficients at various powers of $\ln\Lambda/\mu$ in terms of the coefficients of the $\beta$-function and (in the case of the field renormalization) of the anomalous dimension. As a test, we have verified that these expressions correctly reproduce the coefficients in the renormalization of ${\cal N}=1$ SQED with $N_f$ flavors in the four-loop approximation for the $\beta$-function and in the three-loop approximation for the anomalous dimension of the matter superfields. Note that in the HD+MSL scheme RGFs of ${\cal N}=1$ supersymmetric theories satisfy the NSVZ relation in all orders \cite{Stepanyantz:2020uke}. As was demonstrated in this paper, the relation between the renormalizations of the coupling constant and of the chiral matter superfields following from the NSVZ equation can be constructed with the help of the equations derived in this paper. Moreover, for various theories these equations can be used for multiloop calculations as a nontrivial test of the correctness. Also in the case of using the dimensional technique with $\Lambda\ne \mu$, where $\Lambda$ is the dimensionful regularization parameter, (see, e.g., \cite{Aleshin:2015qqc,Aleshin:2016rrr,Aleshin:2019yqj}) it would be interesting to compare the coefficients at powers of $1/\varepsilon$ and $\ln\Lambda/\mu$. Possibly, the equations obtained in this paper can be useful for this purpose.

\section*{Acknowledgments}
\hspace*{\parindent}

K.S. is very grateful to A.L.Kataev and D.I.Kazakov for indicating the problem of comparing the coefficients at powers of $1/\varepsilon$ and $\ln\Lambda/\mu$, which was one of the motivations for this research, and valuable discussions.

This work was supported by Foundation for Advancement of Theoretical Physics and Mathematics ``BASIS'', grant  No. 21-2-2-25-1 (N.M.) and by Russian Scientific Foundation, grant No. 21-12-00129 (K.S.).

\appendix

\section*{Appendix}

\section{The coefficients in the expansions of $\ln\alpha/\alpha_0$ and $(\alpha/\alpha_0)^S$.}
\hspace*{\parindent}\label{Appendix_Powers}

Repeating the reasoning of Section~\ref{Section_B_nk} in the HD+MSL scheme it is possible to write the coefficients in the expressions

\begin{eqnarray}\label{Ln_Expansion}
&&\ \ln \frac{\alpha}{\alpha_0} \equiv \sum\limits_{n=1}^\infty \alpha_0^n \sum\limits_{m=1}^{n} \widetilde B_{n,m} \ln^m\frac{\Lambda}{\mu}\qquad\quad \text{and}\qquad\\
\label{AlphaS_Expansion}
&& \Big(\frac{\alpha}{\alpha_0}\Big)^S \equiv 1 + \sum\limits_{n=1}^\infty \alpha_0^n \sum\limits_{m=1}^{n} B_{n,m}^{(S)} \ln^m\frac{\Lambda}{\mu}
\end{eqnarray}

\noindent
in terms of $\beta_n$. Note that the case considered in Section~\ref{Section_B_nk} corresponds to $S=-1$,

\begin{equation}\label{B_Minus_1}
B_{n,m}^{(-1)} = B_{n,m}; \qquad B_{p,p}^{(-1)} = 0,\qquad p\ge 2.
\end{equation}

First, we differentiate the expressions (\ref{Ln_Expansion}) and (\ref{AlphaS_Expansion}) with respect to $\ln\mu$ at a fixed value of the bare coupling constant $\alpha_0$ and substitute the perturbative series for the $\beta$-function into the resulting expressions,

\begin{eqnarray}
&& \frac{d}{d\ln\mu} \ln \frac{\alpha}{\alpha_0} = \frac{\beta(\alpha)}{\alpha} = \sum\limits_{k=1}^\infty \beta_k \alpha^k; \\
&& \frac{d}{d\ln\mu} \Big(\frac{\alpha}{\alpha_0}\Big)^S = (\alpha_0)^{-S} S \alpha^{S-1} \beta(\alpha) = (\alpha_0)^{-S} S \sum\limits_{k=1}^\infty \beta_k \alpha^{k+S}.
\end{eqnarray}

\noindent
Then, we differentiate these equations $(p-1)$ times with respect to $\ln\mu$ and, after that, set $\mu=\Lambda$,

\begin{eqnarray}
&& \sum\limits_{k=1}^\infty \beta_k\, \frac{d^{p-1} \alpha^{k}}{d \ln\mu^{p-1}}\bigg|_{\mu=\Lambda}
= \frac{d^p}{d\ln\mu^p} \sum\limits_{n=1}^\infty \alpha_0^n \sum\limits_{m=1}^{n} \widetilde B_{n,m} \ln^m\frac{\Lambda}{\mu}\bigg|_{\mu=\Lambda}; \\
\vphantom{1}\nonumber\\
&& (\alpha_0)^{-S} S \sum\limits_{k=1}^\infty \beta_k\, \frac{d^{p-1} \alpha^{k+S}}{d \ln\mu^{p-1}}\bigg|_{\mu=\Lambda}
= \frac{d^p}{d\ln\mu^p} \sum\limits_{n=1}^\infty \alpha_0^n \sum\limits_{m=1}^{n} B_{n,m}^{(S)} \ln^m\frac{\Lambda}{\mu}\bigg|_{\mu=\Lambda}.\qquad
\end{eqnarray}

\noindent
Calculating the derivatives (in the left hand side for this purpose we use Eq.~(\ref{MainDifferentiations})) gives the equations

\begin{eqnarray}
&&\hspace*{-5mm} \sum\limits_{k=1}^\infty \beta_k  \sum\limits_{k_1=1}^\infty k \beta_{k_1} \sum\limits_{k_2=1}^\infty (k+k_1) \beta_{k_2} \sum\limits_{k_3=1}^\infty (k+k_1+k_2)\beta_{k_3} \dots
\sum\limits_{k_{p-1}=1}^\infty (k+k_1+k_2+\dots+k_{p-2}) \qquad\nonumber\\
&&\hspace*{-5mm} \times \beta_{k_{p-1}} \alpha_0^{k+k_1+k_2+\dots+k_{p-1}}= (-1)^{p} p! \sum\limits_{n=0}^\infty  \alpha_0^{n+p} \widetilde B_{n+p,\,p}; \\
&&\hspace*{-5mm}  S \sum\limits_{k=1}^\infty \beta_k  \sum\limits_{k_1=1}^\infty (k+S)\beta_{k_1} \sum\limits_{k_2=1}^\infty (k+S+k_1) \beta_{k_2} \sum\limits_{k_3=1}^\infty (k+S+k_1+k_2)\beta_{k_3} \dots
\sum\limits_{k_{p-1}=1}^\infty (k+S+k_1 \qquad\nonumber\\
&&\hspace*{-5mm} +k_2+\dots+k_{p-2}) \beta_{k_{p-1}} \alpha_0^{k+k_1+k_2+\dots+k_{p-1}}  = (-1)^{p} p! \sum\limits_{n=0}^\infty  \alpha_0^{n+p} B_{n+p,\,p}^{(S)}.
\end{eqnarray}

\noindent
Equating the coefficients at the same powers of the coupling constant $\alpha_0$ we obtain the required expressions for the coefficients in Eqs.~(\ref{Ln_Expansion}) and (\ref{AlphaS_Expansion}),

\begin{eqnarray}
\label{Ln_Coefficients}
&&\hspace*{-5mm} \widetilde B_{n+p,\,p} = (-1)^{p}\frac{1}{p!}\,\overset{n+1}{\underset{k=1}{\sum}}\, \beta_{k} \sum_{k_1} k \beta_{k_1} \sum_{k_2}(k+k_1) \beta_{k_2}\sum_{k_3} (k+k_1+k_2)\beta_{k_3}\dots
\nonumber\\
&&\hspace*{-5mm} \times \sum_{k_{p-1}}(k+k_1+k_2+\dots+k_{p-2})\beta_{k_{p-1}}\bigg|_{\substack{k_1+k_2+\dots +k_{p-1}=n+p-k}};\\
\vphantom{1}\nonumber\\
\label{AlphsS_Coefficients}
&&\hspace*{-5mm} B_{n+p,\,p}^{(S)} = S (-1)^{p}\frac{1}{p!}\,\overset{n+1}{\underset{k=1}{\sum}}\, \beta_{k} \sum_{k_1} (k+S) \beta_{k_1} \sum_{k_2}(k+S+k_1) \beta_{k_2}\sum_{k_3} (k+S+k_1+k_2)\beta_{k_3}\dots
\nonumber\\
&&\hspace*{-5mm} \times \sum_{k_{p-1}}(k+S+k_1+k_2+\dots+k_{p-2})\beta_{k_{p-1}}\bigg|_{\substack{k_1+k_2+\dots +k_{p-1}=n+p-k}},
\end{eqnarray}

\noindent
where $n\ge 0$ and $p\ge 1$. Certainly, for $S=-1$ Eq.~(\ref{AlphsS_Coefficients}) gives the expressions (\ref{Bk1=bk}) for $p=1$ and (\ref{GeneralAnaliticalResult}) for $p\ge 2$. Note that from Eq.~(\ref{AlphsS_Coefficients}) we obtain

\begin{equation}
B_{p,p}^{(S)} = \frac{(-1)^p}{p!} S(S+1)\ldots (S+p-1)\, \beta_1^p,
\end{equation}

\noindent
so that $B_{p,p}^{(-1)} = 0$ for all $p\ge 2$.


\begin{thebibliography}{100}

\bibitem{Bogolyubov:1980nc}
N.~N.~Bogolyubov and D.~V.~Shirkov,
``Introduction To The Theory Of Quantized Fields,''
Intersci. Monogr. Phys. Astron. \textbf{3} (1959), 1.

\bibitem{'tHooft:1972fi}
G.~'t Hooft and M.~J.~G.~Veltman,
Nucl. Phys. B \textbf{44} (1972), 189.

\bibitem{Bollini:1972ui}
C.~G.~Bollini and J.~J.~Giambiagi,
Nuovo Cim. B \textbf{12} (1972), 20.

\bibitem{Ashmore:1972uj}
J.~F.~Ashmore,
Lett. Nuovo Cim.  \textbf{4} (1972), 289.

\bibitem{Cicuta:1972jf}
G.~M.~Cicuta and E.~Montaldi,
Lett. Nuovo Cim. \textbf{4} (1972), 329.

\bibitem{tHooft:1973mfk}
G.~'t Hooft,
Nucl. Phys. B \textbf{61} (1973), 455.

\bibitem{Kazakov:1987jp}
D.~I.~Kazakov,
Theor. Math. Phys. \textbf{75} (1988), 440.

\bibitem{Kazakov:2016wrp}
D.~I.~Kazakov and D.~E.~Vlasenko,
Phys. Rev. D \textbf{95} (2017) no.4, 045006.

\bibitem{Borlakov:2016mwp}
A.~T.~Borlakov, D.~I.~Kazakov, D.~M.~Tolkachev and D.~E.~Vlasenko,
JHEP \textbf{12} (2016), 154.

\bibitem{Kazakov:2019wce}
D.~I.~Kazakov,
Phys. Lett. B \textbf{797} (2019), 134801.

\bibitem{Solodukhin:2020vuw}
S.~N.~Solodukhin,
Nucl. Phys. B \textbf{962} (2021), 115246.

\bibitem{Collins:1984xc}
J.~C.~Collins,
``Renormalization: An Introduction to Renormalization, The Renormalization Group, and the Operator Product Expansion,''
Cambridge University Press, 1986.

\bibitem{Gnendiger:2017pys}
C.~Gnendiger, A.~Signer, D.~St\"ockinger, A.~Broggio, A.~L.~Cherchiglia, F.~Driencourt-Mangin, A.~R.~Fazio, B.~Hiller, P.~Mastrolia and T.~Peraro, \textit{et al.}
Eur. Phys. J. C \textbf{77} (2017) no.7, 471.

\bibitem{Delbourgo:1974az}
R.~Delbourgo and V.~B.~Prasad,
J. Phys. G \textbf{1} (1975), 377.

\bibitem{Siegel:1979wq}
W.~Siegel,
Phys. Lett. B \textbf{84} (1979), 193.

\bibitem{Avdeev:1981vf}
L.~V.~Avdeev, G.~A.~Chochia and A.~A.~Vladimirov,
Phys. Lett. B \textbf{105} (1981), 272.

\bibitem{Avdeev:1982xy}
L.~V.~Avdeev and A.~A.~Vladimirov,
Nucl. Phys. B \textbf{219} (1983), 262.

\bibitem{Avdeev:1982np}
L.~V.~Avdeev,
Phys. Lett. B \textbf{117} (1982), 317.

\bibitem{Velizhanin:2008rw}
V.~N.~Velizhanin,
Nucl. Phys. B \textbf{818} (2009), 95.

\bibitem{Slavnov:1971aw}
A.~A.~Slavnov,
Nucl. Phys. B \textbf{31} (1971), 301.

\bibitem{Slavnov:1972sq}
A.~A.~Slavnov,
Theor. Math. Phys. \textbf{13} (1972), 1064
[Teor. Mat. Fiz. \textbf{13} (1972), 174].

\bibitem{Krivoshchekov:1978xg}
V.~K.~Krivoshchekov,
Theor. Math. Phys. \textbf{36} (1978), 745
[Teor. Mat. Fiz. \textbf{36} (1978), 291].

\bibitem{West:1985jx}
P.~C.~West,
Nucl. Phys. B \textbf{268} (1986), 113.

\bibitem{Buchbinder:2015eva}
I.~L.~Buchbinder, N.~G.~Pletnev and K.~V.~Stepanyantz,
Phys. Lett. B \textbf{751} (2015), 434.

\bibitem{Shirokov:2022jyd}
I.~Shirokov and K.~Stepanyantz,
JHEP \textbf{04} (2022), 108.

\bibitem{Kuzmichev:2021lqa}
M.~Kuzmichev, N.~Meshcheriakov, S.~Novgorodtsev, V.~Shatalova, I.~Shirokov and K.~Stepanyantz,
Eur. Phys. J. C \textbf{82} (2022) no.1, 69.

\bibitem{Kuzmichev:2021yjo}
M.~Kuzmichev, N.~Meshcheriakov, S.~Novgorodtsev, I.~Shirokov and K.~Stepanyantz,
Phys. Rev. D \textbf{104} (2021) no.2, 025008.

\bibitem{Aleshin:2020gec}
S.~S.~Aleshin, \textit{et al.}
Nucl. Phys. B \textbf{956} (2020), 115020.

\bibitem{Kuzmichev:2019ywn}
M.~D.~Kuzmichev, N.~P.~Meshcheriakov, S.~V.~Novgorodtsev, I.~E.~Shirokov and K.~V.~Stepanyantz,
Eur. Phys. J. C \textbf{79} (2019) no.9, 809.

\bibitem{Stepanyantz:2019lyo}
K.~Stepanyantz,
Proceedings of the Steklov Institute of Mathematics, \textbf{309} (2020), 284.

\bibitem{Kazantsev:2018kjx}
A.~E.~Kazantsev, M.~D.~Kuzmichev, N.~P.~Meshcheriakov, S.~V.~Novgorodtsev, I.~E.~Shirokov, M.~B.~Skoptsov and K.~V.~Stepanyantz,
JHEP \textbf{06} (2018), 020.

\bibitem{Kazantsev:2018nbl}
A.~E.~Kazantsev, V.~Y.~Shakhmanov and K.~V.~Stepanyantz,
JHEP \textbf{04} (2018), 130.

\bibitem{Shakhmanov:2017soc}
V.~Y.~Shakhmanov and K.~V.~Stepanyantz,
Nucl. Phys. B \textbf{920} (2017), 345.

\bibitem{Novikov:1983uc}
V.~A.~Novikov, M.~A.~Shifman, A.~I.~Vainshtein and V.~I.~Zakharov,
Nucl. Phys. B \textbf{229} (1983), 381.

\bibitem{Jones:1983ip}
D.~R.~T.~Jones,
Phys. Lett. \textbf{123B} (1983), 45.

\bibitem{Novikov:1985rd}
V.~A.~Novikov, M.~A.~Shifman, A.~I.~Vainshtein and V.~I.~Zakharov,
Phys. Lett. \textbf{166B}(1986), 329
[Sov. J. Nucl. Phys. \textbf{43} (1986), 294]
[Yad. Fiz. \textbf{43} (1986), 459].

\bibitem{Shifman:1986zi}
M.~A.~Shifman and A.~I.~Vainshtein,
Nucl. Phys. B \textbf{277} (1986), 456
[Sov. Phys. JETP \textbf{64} (1986), 428]
[Zh. Eksp. Teor. Fiz. \textbf{91} (1986), 723].

\bibitem{Stepanyantz:2016gtk}
K.~V.~Stepanyantz,
Nucl. Phys. B \textbf{909} (2016), 316.

\bibitem{Stepanyantz:2019ihw}
K.~V.~Stepanyantz,
JHEP \textbf{10} (2019), 011.

\bibitem{Stepanyantz:2019lfm}
K.~V.~Stepanyantz,
JHEP \textbf{01} (2020), 192.

\bibitem{Stepanyantz:2020uke}
K.~Stepanyantz,
Eur. Phys. J. C \textbf{80} (2020) no.10, 911.

\bibitem{Kataev:2013eta}
A.~L.~Kataev and K.~V.~Stepanyantz,
Nucl. Phys. B \textbf{875} (2013), 459.

\bibitem{Shakhmanov:2017wji}
V.~Y.~Shakhmanov and K.~V.~Stepanyantz,
Phys. Lett. B \textbf{776} (2018), 417.

\bibitem{Stepanyantz:2017sqg}
K.~V.~Stepanyantz,
Bled Workshops Phys. \textbf{18} (2017) no.2, 197.

\bibitem{Slavnov:1977zf}
A.~A.~Slavnov,
Theor. Math. Phys. \textbf{33} (1977), 977
[Teor. Mat. Fiz. \textbf{33} (1977), 210].

\bibitem{Aleshin:2016yvj}
S.~S.~Aleshin, A.~E.~Kazantsev, M.~B.~Skoptsov and K.~V.~Stepanyantz,
JHEP \textbf{05} (2016), 014.

\bibitem{Kazantsev:2017fdc}
A.~E.~Kazantsev, M.~B.~Skoptsov and K.~V.~Stepanyantz,
Mod. Phys. Lett. A \textbf{32} (2017) no.36, 1750194.

\bibitem{Vainshtein:1986ja}
A.~I.~Vainshtein, V.~I.~Zakharov and M.~A.~Shifman,
JETP Lett. \textbf{42} (1985), 224
[Pisma Zh. Eksp. Teor. Fiz. \textbf{42} (1985), 182].

\bibitem{Shifman:1985fi}
M.~A.~Shifman, A.~I.~Vainshtein and V.~I.~Zakharov,
Phys. Lett. B \textbf{166} (1986), 334.

\bibitem{Korneev:2021zdz}
  D.~Korneev, D.~Plotnikov, K.~Stepanyantz and N.~Tereshina,
  JHEP \textbf{10} (2021), 046.

\bibitem{Aleshin:2015qqc}
S.~S.~Aleshin, A.~L.~Kataev and K.~V.~Stepanyantz,
JETP Lett. \textbf{103} (2016) no.2, 77.

\bibitem{Aleshin:2016rrr}
S.~S.~Aleshin, I.~O.~Goriachuk, A.~L.~Kataev and K.~V.~Stepanyantz,
Phys. Lett. B \textbf{764} (2017), 222.

\bibitem{Aleshin:2019yqj}
S.~S.~Aleshin, A.~L.~Kataev and K.~V.~Stepanyantz,
JHEP \textbf{03} (2019), 196.


\end{thebibliography}
\end{document}